\documentclass[10pt, conference, letterpaper]{IEEEtran}

\usepackage[english]{babel}
\usepackage{multirow}
\usepackage{textcomp}
\usepackage{booktabs} 
\usepackage{subcaption}
\usepackage{microtype}
\usepackage{pgfplotstable} 
\usepackage{hyphenat}
\usepackage[ruled,vlined]{algorithm2e}
\usepackage{setspace} %

\usepackage[cmex10]{amsmath}
\usepackage{amsfonts}
\usepackage{amssymb}
\usepackage{cases}
\pgfplotsset{compat=newest}
\pgfplotstableset{fixed zerofill, sci zerofill, precision=2}
\pgfkeys{/pgf/number format/.cd,std=-5:4,precision=2}
\pgfkeys{/pgf/number format/sci e}
\pgfkeys{/pgf/number format/sci generic =  {mantissa sep=e{#1}}}
\pgfkeys{/pgf/number format/sci precision=2}

\SetKwProg{Fn}{Function}{}{end}
\SetKwProg{Str}{Structure}{}{end}

\newcommand{\figspace}{-0.3em} %

\newcommand{\tename}{HCTE\xspace}
\newcommand{\fulltename}{\emph{Hop-by-Hop Congestion-Aware Traffic Engineering}\xspace}

\usepackage{setspace} %

\begin{document}
\title{
The Case for Hop-by-Hop Traffic Engineering
}

\author{\IEEEauthorblockN{Klaus Schneider, Beichuan Zhang}
\IEEEauthorblockA{University of Arizona\\
Email: \{klaus\textbar bzhang\}@cs.arizona.edu}
\and
\IEEEauthorblockN{Van Sy Mai, Lotfi Benmohamed}
\IEEEauthorblockA{National Institute of Standards and Technology\\
Email: \{vansy.mai\textbar lotfi.benmohamed\}@nist.gov}
}

\maketitle

\begin{abstract}

State-of-the-art Internet traffic engineering uses source-based explicit routing via MPLS or Segment Routing. 
Though widely adopted in practice, 
source routing can face certain inefficiencies and operational issues, caused by its use of bandwidth reservations. 

In this work, we make the case for Hop-by-Hop (HBH) Traffic Engineering: 
splitting traffic among nexthops at every router, 
rather than splitting traffic among paths only at edge routers.
We show that HBH traffic engineering can achieve the original goals of MPLS (i.e., efficient use of network resources), with a much simpler design that does not need bandwidth reservations or predictions of traffic demand. 

We implement a prototype in the ns-3 network simulator, to investigate the cost imposed by 1) the restricted path choice of loop-free HBH multipath routing, and 2) the distributed decisions of each router, based on its local network view.
We show that the former is more important than the latter,
but that, other than a few outliers, our design shows a performance (= aggregate user utility) close to the theoretical optimum.

\end{abstract}

\section{Introduction}

While conventional routing protocols (like OSPF \cite{ospf_v2_rfc2328} or IS-IS \cite{isis_rfc1142}) send packets on the shortest path from source to destination, there are many benefits of using multiple \emph{non-shortest paths} \cite{rexford2008toward}, most importantly \emph{failure protection:} routing around link/node failures, and \emph{traffic engineering:} distributing the traffic load to avoid congestion and improve user satisfaction.

A first approach to traffic engineering was to \emph{tune the link weights} of traditional routing protocols.
The weights of highly loaded links are increased, so they are less likely to be used, and their load decreases.
While link-weight tuning is simple to implement, it can lead to abrupt shifts in traffic and hard-to-predict network-wide side-effects.
A famous example is the early ARPANET, where dynamic changes to link weights have lead to instability and routing oscillations \cite{revised1989}.

A way to solve these issues, is to use \emph{source routing}, e.g., MPLS \cite{mpls_rfc3031} or Segment Routing (SR) \cite{sr_rfc8402}:
letting the source (ingress) node determine the entire path (or tunnel) towards the destination. 
If a link becomes overloaded, traffic is shifted to a tunnel that avoids this link, leading to much more predictable outcomes than link-weight tuning.
Though frequently used in practice, source routing does have certain downsides (see Section \ref{sec:history_source}): 
1) It is unclear which tunnels should be chosen to achieve a good trade-off between shorter paths and failure resilience.
2) Tunnels often have large bandwidth reservations and can only be moved as an atomic unit, which leads to sub-optimal path choices.
3) Online adaptation of tunnel reservation size (called ``Auto-Bandwidth'') does not directly consider congestion in the network, which can lead to inefficiencies.

\begin{figure}
\footnotesize
\begin{subfigure}{0.4\linewidth}
\begin{tabular}{@{}llcr@{}}
\toprule
\textbf{Dest.} &  \textbf{NH} & \textbf{\hspace*{-.3em}Cost\hspace*{-.4em}} & \textbf{Split} \\ 
\midrule
LA 											& SV	& 2 & 100\%  \\
\midrule
\multirow{2}{0em}{DV}   & DV	& 1 & 100\%  \\
									    	& SV	& 2 &   0\%  \\
\midrule
\multirow{2}{0em}{NYC}  & DV	& 5 &  60\%  \\
									    	& SV	& 6 &  40\%  \\
\midrule
\ldots \\ 
\bottomrule
\end{tabular}
\caption{FIB for \textbf{Seattle}}
\label{fig:fib_example_a}
\end{subfigure}
\begin{subfigure}{.58\linewidth}
\includegraphics[width=\linewidth]{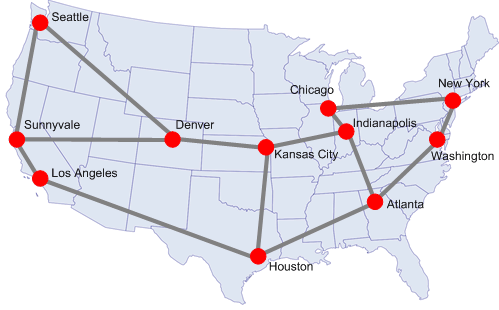}
\caption{Abilene Topology}
\label{fig:fib_example_b}
\end{subfigure}

\caption{Example of HBH Traffic Engineering. Depending on the destination, Seattle uses either 1 or 2 nexthops. The cost denotes the shortest path (in hops) of using a certain nexthop. The split ratio can be adapted to react to link failures or congestion.
\vspace{\figspace}}
\label{fig:fib_example}
\end{figure}

In this work, we adopt Hop-by-Hop Traffic Engineering (HBH TE) to avoid the issues of source routing, with the goal to create an overall much simpler design. 
The basic idea of HBH TE is as follows:
1) A multipath routing protocol equips each router with a list of loop-free nexthops per destination, ranked by their routing distance (see Figure \ref{fig:fib_example_a}). 
This list and ranking is fixed, i.e., will not change based on traffic load.
2) Each router maintains a split ratio (or ``forwarding percentage'') for each nexthop in the FIB. 
Routers will adapt this split ratio based on the traffic load to meet the goals of traffic engineering. 

The main two questions are which nexthops to choose and how to set the split ratio? 
We leave the first question to prior work (discussed in Section \ref{sec:design_mp_routing}).
For the second, we use the goal of what we call ``latency-optimality'' (Section \ref{sec:optimal}): 
packets should take the shortest path (by propagation delay) unless the shortest path's capacity is insufficient to meet traffic demand; then packets should take longer paths to maximize application throughput. 
Surprisingly, most existing work on traffic engineering chooses a different goal (Section \ref{sec:related_work}): minimizing the maximum link utilization or cost. 
These different goals lead to quite different designs. For example, while most existing work periodically measures link utilization, we only need to measure link \emph{congestion}, that is, the state when link utilization approaches capacity.

Two obvious disadvantages of HBH TE compared to source routing TE need to be addressed. 
First, source routing can choose arbitrary paths, while HBH TE's paths are restricted by the requirement of loop-free hop-by-hop routing without per-packet state; that is, by which nexthops can be put in the FIB without risking loops.
We show that this restricted path choice does indeed lead to sub-optimal traffic distribution, but only in rare cases, and that the omitted paths are many times longer than the shortest path (Section \ref{sec:ev_geant}). 
Second, routers make decisions based on a local view of the network, as opposed to the global view achieved by source routing with a centralized controller. 
We show that, perhaps surprisingly, the lack of global knowledge makes little difference. 
By avoiding the granularity problem of bandwidth reservations together with path probing, our design comes close to the theoretical optimum (Section \ref{sec:ev_geant}).
Divergence from the optimum is caused much more by the limited path choice than by local decision making.

Our specific design, \fulltename (\tename), works as follows: 
it starts forwarding packets on the shortest path, splits traffic only when these paths become congested (detected by ECN congestion marks), probes the congestion level of new paths before use (and only adds paths with a lower congestion level than the current), 
adds new paths gradually,
and equalizes the congestion price (based on queue occupancy) in equilibrium (see Section \ref{sec:design}).

Earlier designs of HBH TE are quite rare (we only found two of them -- see Section \ref{sec:related_work}) and not widely deployed yet. 
Hence, with this work, we want to make a renewed case for Hop-by-Hop Traffic Engineering. 
The main difference over the two existing works is our different definition of optimality (Section \ref{sec:optimal}), and a novel way of probing the path congestion price to adapt the split ratio.

Compared to source routing, a strong selling point is that \tename aims to be vastly \emph{simpler:}
It does not require 1) estimating traffic matrices, 2) pre-selecting a good set of paths (only a good set of potential nexthops), or 3) reserving bandwidth for subsets of traffic.
It aims to work out of the box for many topologies, link capacities, and traffic demand patterns.
We show that it comes close to the theoretically optimal performance in terms of average completion time and user utility.
Moreover, we show that loop-free HBH routing mostly finds the same paths as optimal arbitrary source routing with a global view of traffic demand and link utilization.
And that the omitted paths are many times longer than the shortest path (Section \ref{sec:evaluation}).

\section{The History of Traffic Engineering}
\label{sec:history}

\subsection{Link Weight Tuning}

Sending packets on the shortest path to the destination is not always the best way to use network resources.
Some of these shortest paths may be congested, while other, longer paths remain underutilized. 
Consider Figure \ref{fig:ecmp_tuning_base}, where Seattle (SE) sends traffic to Kansas City (KC) over the shortest path SE$\rightarrow$DV$\rightarrow$KC.
If link (SE, DV) or link (DV, KC) are congested, one would want to reroute traffic over the paths SE$\rightarrow$SV$\rightarrow$DV$\rightarrow$KC, or respectively SE$\rightarrow$SV$\rightarrow$LA$\rightarrow$HOU$\rightarrow$KC.

One can achieve this sort of \emph{traffic engineering} by tuning the link-weights of conventional routing protocols (like OSPF or IS-IS) \cite{fortz2000internet, fortz2002traffic, iannaccone2004feasibility}, often combined with Equal-Cost Multipath (ECMP). 
For example, changing the weights of (SE, DV) and (DV, KC) to 2 allows ECMP to use all three paths simultaneously (Figure \ref{fig:ecmp_tuning_manip}).
 
However, using ECMP for paths of different length (= physical distance or propagation delay), comes with undesirable side-effects:
\begin{figure}
\footnotesize
\centering
\begin{subfigure}{.315\linewidth}
\includegraphics[width=\linewidth]{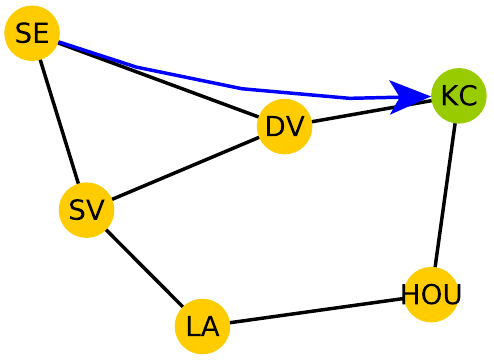}
\caption{} 
\label{fig:ecmp_tuning_base}
\end{subfigure}
\hspace*{2em}
\begin{subfigure}{.315\linewidth}
\includegraphics[width=\linewidth]{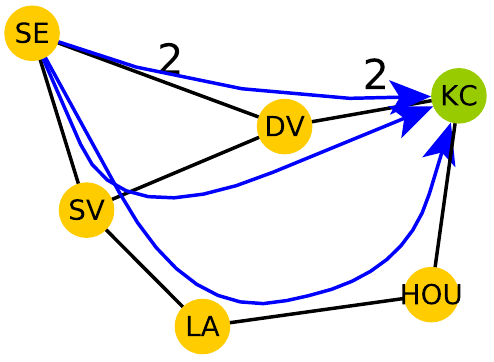}
\caption{}
\label{fig:ecmp_tuning_manip}
\end{subfigure}
\caption{Example of ECMP + Link weight tuning. a) Base topology that allows only one equal-cost path. b) Topology with changed link weights to create three equal-cost paths. 
\vspace{\figspace}}
\label{fig:ecmp_tuning}
\end{figure}
The only way ECMP can use these paths is to split traffic on them \emph{evenly}, i.e., every path receives about 33\% of traffic (Figure \ref{fig:ecmp_tuning_manip}).
As long as link utilization is below capacity, this is wasteful: some traffic that could have used the shortest path (SE$\rightarrow$DV$\rightarrow$KC) is needlessly sent over a longer path.
A better approach would be to keep traffic on the shortest path until demand exceeds its capacity, and only then \emph{gradually} switch to the longer path (see Section \ref{sec:optimal}).
However, ECMP + link weight tuning cannot achieve this, as it mandates an even split between all equal-cost paths.

ECMP + link-weight tuning has further problems: 1) in larger topologies, it is hard to find link weights that create the desired ECMP paths, 2) an equal split may not be what's needed to create equal link utilization (e.g. when links differ in their capacity), and 3) changing link-weights can have hard-to-predict network-wide side-effects, i.e., causing congestion in other parts of the network.

\subsection{Source-based explicit routing}
\label{sec:history_source}

A solution to these issues is MPLS/Segment Routing (SR) traffic engineering, or more generally, \emph{source-based explicit routing} through tunnels between network edge routers. 
In source routing, ingress routers set up a number of tunnels (label-switched paths -- LSPs) towards any given egress router. 
Each tunnel will \emph{reserve} a specific amount of bandwidth and can be set up automatically via the Constrained Shortest Path First (CSPF) algorithm, i.e., finding the shortest path that will also satisfy the bandwidth requirement. 
Hence, source-based routing solves the problems of link-weight tuning: It can arbitrarily split up traffic over multiple non-shortests paths/tunnels, without causing network-wide side-effects. 

Segment Routing (SR) \cite{sr_rfc8402} is an important update to MPLS (using RSVP-TE \cite{rsvp_te_rfc}), as it removes the need for pre-establishing paths and maintaining state inside the network. 
SR puts all the state needed to route a packet into labels attached to the packet itself. %
However, for our discussion, the similarities between MPLS and SR are more important than the differences: 
forwarding decisions are made only at the network edge, as 
both protocols send packets on explicit paths/tunnels between ingress and egress routers. Moreover, both use LSP bandwidth reservations as means of traffic engineering.

Even though explicit source routing avoids the problems of link-weight tuning, it faces a number of operational problems itself
\cite{mpls_autobw} \cite{mpls_autobw2}:

\paragraph{Manual choice of LSPs (number \& which ones)}

As Jennifer Rexford points out \cite{rexford_rethinking}, there is no systematic way of determining which paths should be used between two endpoints. 
Using the k shortest paths may not give enough disjointness to deal with failure or congestion.
Using the k maximally disjoint paths may lead to paths that are unnecessarily long. 
As Rexford puts it, there is an ``art'' involved in finding k short paths that are also reasonably disjoint.

\paragraph{Granularity: Large tunnels don’t fit into small pipes}

LSPs can only be moved as an atomic unit. If an LSP's bandwidth reservation is a large fraction of the bottleneck link capacity, it might not fit on the optimal path, or not fit on any path at all \cite{mpls_autobw}.
In case the LSP does not fit, traffic will still flow, but not be accounted by MPLS bandwidth reservations, which can lead to problems like congestion later on.

\begin{figure}
\centering
\includegraphics[width=.95\linewidth]{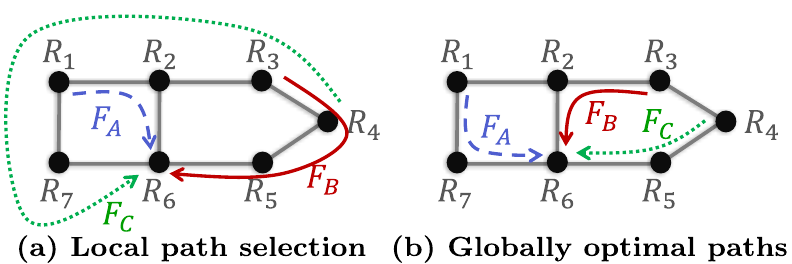}
\caption{Inefficient routing in MPLS TE \cite{hong2013swan}
\vspace{\figspace}
}
\label{fig:hong_mpls}
\end{figure}

Hong et al. \cite{hong2013swan} give an illustrative example of the granularity problem (Figure \ref{fig:hong_mpls}). 
Assuming that links have equal capacity and can transfer exactly one flow, flows $F_A$, $F_B$, and $F_C$ are subsequently added using CSPF. 
This greedy allocation leads to undesirable results: 
$F_B$ and $F_C$ take much longer paths then necessary (Figure \ref{fig:hong_mpls}a).
Hong et al. attribute this problem to MPLS' ``local, greedy resource allocation'', which they solve via an SDN controller with global knowledge. 

But the problem could just as well be solved locally, if the \emph{granularity} constraint was removed.
$R_4$ (locally) knows that the path over $R_4\rightarrow R_2\rightarrow R_1 \rightarrow R_7\rightarrow R_6$ is suboptimal and only chooses it, because the other paths are already fully utilized. 
Without granularity constraints, $R_4$ could shift over a fraction of the traffic of $F_C$ to path $R_5 \rightarrow R_6$, which will then cause $R_3$ to shift some of $F_B$ to $R_2 \rightarrow R_6$, which will cause $R_1$ to shift some of $F_A$ to $R_7 \rightarrow R_6$, slowly converging to an equilibrium of the globally optimal solution.
This is indeed what happens in \tename (see Section \ref{sec:evaluation}).

\begin{figure}
\centering
\includegraphics[width=.9\linewidth]{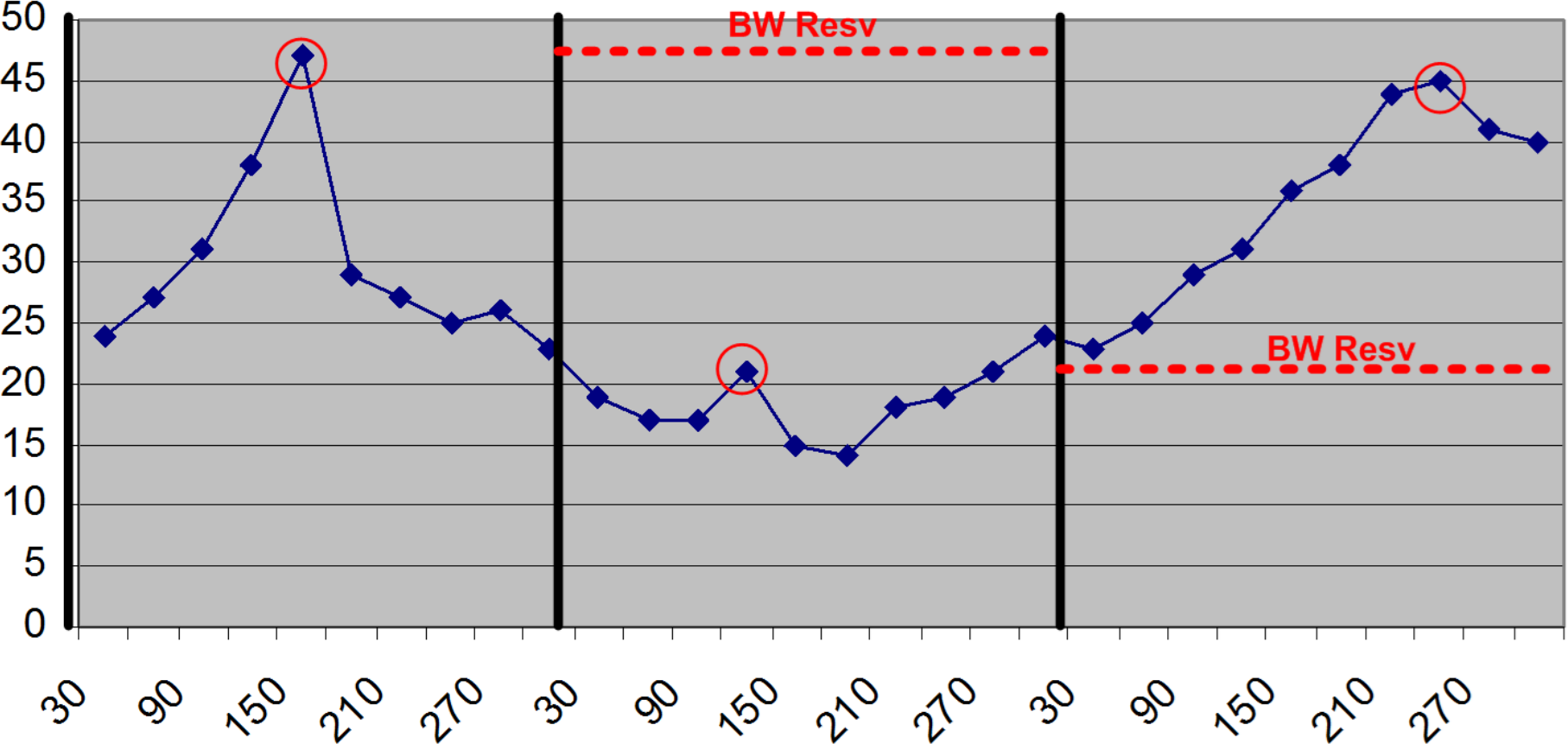}
\caption{When Auto-Bandwidth Doesn't Work Well \cite{mpls_autobw}
\vspace{\figspace}
}
\label{fig:auto_bw_problems}
\end{figure}

\paragraph{Auto-BW does not measure congestion directly}
There are two ways to adjust an LSP's bandwidth assignment: 1) Offline calculation and 2) ``Auto-Bandwidth'', meaning an online adjustment based on the observed traffic load. \cite{mpls_autobw, mpls_autobw2}. Since offline calculation usually acts on much slower timescales, we focus on auto-bandwidth. 
Auto-BW measures the traffic load over one interval (in the order of minutes), and uses the maximum observed traffic to adjusts the reservation for the next interval (see Figure \ref{fig:auto_bw_problems}).
This allows Auto-BW to use shorter latency paths first, and also react to bursts of traffic in near real-time.

However, if the traffic fluctuates quickly relative to the adjustment interval, Auto-BW may over-reserve or under-reserve bandwidth. See the example in Figure \ref{fig:auto_bw_problems}, where the adjustment interval is 300s.
Over-reserving bandwidth is a problem since it might block other LSPs from using certain paths that have sufficient capacity for them. 
Under-reserving can lead to even worse outcomes, since it can create congestion on paths that are not seen by the Auto-BW mechanism! \cite{mpls_autobw}. 
Creating congestion will have TCP endpoints slow down, reducing the traffic demand seen by Auto-BW, which then sees no need to increase the LSPs size. 
This is a vital flaw of a system that is not congestion-aware.

In summary, there are many conditions where SR/MPLS TE does not lead to optimal results and either human operator intervention or proprietary solutions are necessary \cite{mpls_autobw}.

\section{What is Optimal?}
\label{sec:optimal}

We adopt the following definition of optimality:
``In optimal routing, traffic from a source-destination pair is split at \emph{strategic points} allowing \emph{gradual shifting} of traffic among alternative paths. [\ldots]
For low input traffic, a source-destination pair tends to use \emph{only one path} (which is the fastest in terms of packet transmission time), 
while for increasing input traffic, additional paths are used to avoid overloading the shortest path.'' \cite{qadir2015exploiting} 
To differentiate from other definitions of optimality, we call this ``latency-optimal routing'' \footnote{Note that this optimization goal also works for link cost choices other than latency. E.g. wireless links can be assigned a cost higher than their propagation delay, to discourage their use. 
However, in most cases, latency is the most sensible choice.}.

Even though latency-optimal routing makes intuitive sense,
most of the traffic literature uses a different definition of optimality:
minimizing the maximum link utilization or link cost \cite{fortz2000internet,mate_elwalid2001,texcp2005walking,xu2007deft,xu2011peft,michael2013optimal,michael2015halo}, called \emph{MinMax} routing.
However, MinMax routing in itself has no benefits for application performance (applications work just as well with links utilized at 60\% than at 40\%), 
and can even hurt delay-sensitive applications by using longer paths than necessary \cite{sharma2011beyond}. 
MinMax routing's fundamental problem is that it models application utility based only on throughput but not on latency, and thus optimizes for maximal headroom with no regards to path length.

Leaving some headroom is indeed valuable in order to deal with traffic spikes. 
As Gvozdiev et al. point out, there is a trade-off between using shorter paths (latency-optimal routing) and leaving maximal headroom (MinMax routing)  \cite{on_low_lat_topos}. 
In practice, one might want to leave a defined amount of headroom (e.g. 10\% of link capacity) and aim for latency-optimal routing with that constraint.
See Section \ref{sec:design} for how we consider this requirement in our design.

\section{Hop-by-Hop Traffic Engineering}

We propose a new way of traffic engineering that considers both the problems of link-weight tuning and source routing, and optimizes for latency rather than headroom.
Our approach, \fulltename (\tename),
makes routing decisions at each \emph{hop:} 
Instead of having edge routers determine the \emph{entire path} from ingress to egress, every router inside the network determines the \emph{nexthop} of a packet, based on the destination address, topology information from the routing protocol, and from probing of potential paths/nexthops. 
It uses \emph{HBH Multipath Routing:} each router is given a set of loop-free nexthops by the routing protocol, and will set a fine-grained split ratio among them (see Figure \ref{fig:fib_example}).  
\tename is also \emph{congestion-aware:} The split ratio is based on the congestion level of the paths following the potential nexthops.

\tename solves the problems of link-weight tuning: 
1) It avoids unpredictable network-wide side-effects, since one router changing its split ratio will not cause other nodes to reroute, as can happen when changing link weights.
2) It provides ``latency-optimal routing'': it starts on the shortest path and only splits when necessary (paths are becoming congested). 
3) It can set non-equal and fine-grained split ratios, using the available link capacity more effectively than ECMP.

\tename also solves the problems of source routing: 
1) It does not have to choose an arbitrary set of paths (e.g. k-shortest), but rather chooses from small number of nexthops at each step, which leads to an exponential number of possible paths, used on demand. 
2) It does not face the same granularity issues: 
Split ratios can be chosen arbitrarily fine-grained. 
For practical purposes, it can still be useful to keep all packets of a given TCP flow on the same path (e.g. via flowlet hashing \cite{sinha2004flowlet}), but these flows are typically much smaller than MPLS/SR LSPs.
3) It is directly congestion-aware, which avoids the problem of over- or under-reserving bandwidth. 
Paths can no longer be blocked by underused LSPs, and traffic will only be split when actual utilization approaches link capacity.

Mainly, \tename aims to be vastly \emph{simpler:}
It does not require 1) estimating traffic matrices, 2) deciding on a good set of paths, or 3) reserving bandwidth for subsets of traffic.
It aims to work out of the box for many topologies, link capacities, and traffic demands, with close to optimal (by our definition) performance.
\tename uses only \emph{one main parameter:} the amount (percentage) of adjustment of the forwarding split ratio per probing interval.
All the other parameters have intuitive default values. The probing interval, for example, should be on the order of the maximum possible round-trip time of the network.

\section{Design Specifics}
\label{sec:design}

Following our earlier definition of ``latency-optimal routing'', we start out by sending all traffic on the shortest path (= lowest propagation delay). 
\tename then has to answer the following questions:

\begin{enumerate}
\item When to start splitting traffic? What triggers traffic to be sent on paths other then the shortest path?
\item Where to split traffic? Which router should split and how to coordinate routers?
\item What determines the equilibrium split ratio?
\end{enumerate}
Splitting traffic away from the shortest path creates a trade-off between using shorter paths and achieving higher throughput through longer paths.
\emph{When to split} is thus intuitively answered: whenever the shortest path capacity is insufficient to meet demand.
It also suggest an \emph{equilibrium split ratio:} as long as demand is not met, try to equalize the congestion price on all paths, which will maximize throughput.
\emph{Where to split} is als crucial, since it is possible to choose a longer path that goes back to the same shared bottleneck link, increasing latency without adding throughput. 
Consider Figure \ref{fig:abilene}, and the shortest path SE$\rightarrow$DV$\rightarrow$KC$\rightarrow$IN where link (KC, IN) is the bottleneck. If SE shifts traffic to its nexthop SV, the second shortest path would be SE$\rightarrow$SV$\rightarrow$DV$\rightarrow$KC$\rightarrow$IN, still using the same bottleneck link!
This problem of \emph{shared bottlenecks} is especially pressing for Hop-by-Hop traffic engineering, since routers don't determine the full path of packets.
Our solution is to only split traffic close enough to the bottleneck link, so that traffic on the longer path can never return to that link.
Hence, a quick summary of \tename is as follows. For each destination in the FIB:

\begin{enumerate}
\item Start on the shortest path.

\item When a link gets congested: \emph{gradually} shift traffic away from this link at the router \emph{directly adjacent,} or if not possible, at the \emph{closest possible} router. 
This avoids the problem of returning to a shared bottleneck. 

\item Before splitting traffic, probe alternative nexthops/paths to make sure they have free capacity.

\item In equilibrium: try to equalize the congestion price of all active paths.

\item If demand falls below link capacity: Gradually shift traffic back to the shortest path.

\end{enumerate}

\subsection{Loop-Free Multipath Routing}
\label{sec:design_mp_routing}

We use a multipath link-state routing protocol to fill the FIB with viable nexthops, sorted by their cost (see Figure \ref{fig:fib_example_a}). 
Cost here is defined as the distance of the shortest path to the destination through this particular nexthop. 
We set the link metric to the propagation delay. 
Since in modern high-speed routers (assuming functional traffic engineering), propagation delay vastly exceeds queuing delay and processing delay, this allows routers to have a basic estimate of the latency of the shortest path through each of their nexthops towards the destination.
This latency can be computed efficiently and locally by running Dijkstra's shortest algorithm not just for the node itself, but also once for every neighbor \cite{source_selectable_2006}. 

The multipath routing protocol should calculate which nexthops to put in the FIB, with the constraint to 1) avoid loops and 2) allow as many and as diverse paths as possible. 
An obvious solution is to use only ``Downward Paths'', that is, nexthops that lead closer to the destination, as measured by the link metric. 
Another option is MARA \cite{mara2009} which creates a directed acyclic graph per destination.
We chose LFID \cite{schneider2019hop}, since it allows a greater number of nexthops.

\subsection{Congestion Price \& Active Queue Management}

We answer the question \emph{When to start splitting traffic?} with ``as soon as links on the shortest path start getting congested''. 
We measure this congestion directly at each router queue via Active Queue Management (AQM). 
The idea of AQM schemes like Random Early Detection (RED) \cite{red_floyd1993random} is to drop or mark packets early to signal TCP endpoints to slow down before the queue buffer overflows. 
This can be viewed as charging a link \emph{price} that is a function of the queuing delay, which in itself is determined by the link utilization. 
For most utilizations the price will be 0, but will sharply increase once utilization approaches link capacity, that is, the time the AQM starts marking/dropping packets. 
When using AQM in the marking mode, it will set a 1-bit ECN mark in the packets, to which TCP will react by reducing the sending rate.
The ratio of marked to unmarked packets can be interpreted as an implicit signal of the congestion price \cite{athuraliya2001rem}.

For our design, we use both the implicit price of ECN marks and an explicit congestion price retrieved via probing. 
Traffic splitting will start once a router receives the first marked packet for a given destination node (ECN packets are carried back on TCP Acknowledgements). 
It will then start probing this destination to gain a more accurate explicit congestion price for each of its nexthops. 
We calculate the congestion price as an exponential moving average (EMA) of packets marked by the AQM per time interval. 
We chose the smoothing factor $\alpha$ to model a simple moving average over an interval of 5 seconds, with the well-known approximation $\alpha =2/(N+1)$ (e.g., N = 5000, with one EMA update per millisecond). 
The exact choice of $\alpha$ is less important, since we are comparing \emph{relative} prices of paths, rather than absolute ones.

Setting the price to ECN marks per second has the benefit of leaving a wide choice of AQM algorithms to use. 
We chose the CoDel AQM \cite{codel_nichols2012} for its simplicity and 0-parameter design, but it is easy to swap out, should a better AQM become available. 

One may want to split traffic earlier than AQMs signal congestion, in order to leave some headroom for traffic spikes. 
We suggest two ways to do so in \tename:
First, one could simply calculate the price based on the current link utilization, using a convex price function that sharply increases at the desired level of headroom (e.g. 90\% of link utilization). 
Second, one can use an AQM scheme with a \emph{virtual queue}, which simulates how many packets would be marked/dropped for a (virtual) link of a smaller capacity \cite{avq_kunniyur2001}. 
Both of these ideas require to add a new 1-bit field to packets that indicates congestion to routers (to start initial probing), but not to TCP endpoints.

\subsection{Probing Paths \& Shifting Traffic}

After receiving the first congestion mark, routers will probe new paths before using them.
Specifically, after the router adjacent to the congested link receives the first congestion mark, it sends out an \emph{initial} probing request packet to all (or a predetermined number of) its nexthops. 
To reduce overhead, this initial probing can be limited in frequency; we limit it to once per 10 seconds. 
The probing packets will travel on the \emph{primary path}, that is, the path with the highest split ratio (by default: the shortest path) to the destination, where they are answered by probing replies in the reverse directions. 
For every link that these probing packets traverse, the AQM will add its current congestion price to the packet. 
Hence the router that sent the original probe will receive the sum of the prices for the primary paths through all of its nexthops. 

After receiving the price for all nexthops, they are sorted in ascending order, first by their congestion price, and on a tie by their routing distance. 
The first in the list is called \emph{minPriceNexthop}, the last \emph{maxPriceNexthop}.
A router will then shift a small percentage (by default we use 0.1\%) of traffic from the maxPriceNexthop towards the minPriceNexthop. 
If there are multiple nexthops with a price of 0 (a common case), the one with the shortest distance is chosen as minPriceNexthop. 
Hence, if all nexthops have a price of 0 (common when demand falls below capacity), traffic shifts back to the shortest path. 

If the minPriceNexthop had not been used earlier, it will now be added to the set of \emph{active} nexthops. We define as active all nexthops with a forwarding percentage above 0\%.

If there is more than one active path/nexthop, all active paths will be probed \emph{periodically}. The probing interval should be set at least as high as the maximum RTT in the network; by default we use 200 ms. 
The split ratio adjustment works as described above, shifting from minPriceNexthop to maxPriceNexthop.
The periodic probing interval (200 ms) together with the adjustment per interval (0.1\%) can be used to calculate the maximal convergence speed to shift one nexthop from 0\% to 100\%: 200 seconds. 
We found this to be an acceptable convergence time (lower than what's typical in MPLS Auto-BW), which leads to very little oscillation of the split ratio (\textless 3\%) in equilibrium state. These 3\% are per active destination node, and having traffic of multiple destinations at one router will smooth traffic out further.
The amount of adjustment per interval can be tuned further to achieve a different trade-off between convergence speed and oscillation in equilibrium.
 
Combining rare (once/10s) initial probing of all nexthops with more frequent periodic probing of active nexthops will \emph{gradually add paths/nexthops} (from short \& low congestion to long \& high congestion) until all of their congestion prices exceed 0, meaning their bottleneck link utilization is at capacity. 
In the equilibrium state all active nexthops will have an \emph{equal congestion price.} 
Nexthops that have a higher congestion price than this equilibrium (e.g. are used by a larger number of sources with different destination) will not be used for this destination. 
This will maximize the aggregate throughput, as new paths with free capacity (or lower congestion) are added, but more congested paths are avoided.

In most cases, probing and splitting will happen only at the router \emph{directly adjacent} to the congested link. 
This router will mark the packet with a 1-bit flag as ``handled'', so that routers further back in the network will ignore the set congestion mark.
One exception is when this router only has a single viable nexthop towards the destination, hence cannot perform any splitting: then it will not mark the packet as handled, and a router further back will do the splitting. 
Only reacting directly adjacent (or as close as possible) to the congested link has a vital benefit: 
it avoids unnecessary splitting over paths that still use this link. 
Consider the Abilene topology in Figure \ref{fig:abilene}.
If link (KC, IN) is congested, splitting only at node KC will ensure that whatever second path is used, it will never include link (KC, IN). 
When splitting at a node further back, say SV, one could end up with the new path SV$\rightarrow$LA$\rightarrow$HOU$\rightarrow$KC$\rightarrow$IN, still of them using the congested link (KC, IN) ! 

The total traffic overhead of probing is very small. 
Probe requests and replies need to include an id of the probing origin (2 Byte), of the destination (2 Byte), and for the congestion price (4 Byte float).
Even combined with L2 and IPv6 headers, the packets will be less than 100 Bytes in size and sent once every 200 ms per \emph{active} destination (probing will stop if a router observes no traffic for a given destination). 
Even if a router sees traffic towards 100 active destination nodes, the total probing traffic will be less than 400 Kilobits/s, negligible compared to modern link capacities.

\subsection{Fairness \& Competing Traffic}

Since the split ratio is determined per destination, often traffic for multiple destinations will compete for the same link capacity. 
For this case, should we enforce some notion of fairness between multiple destinations? Make sure that every destination gets at least some of the capacity?

The answer is no. We think it is best to leave fairness up to the endpoints. 
TCP endpoints already enforce a fair sharing of bandwidth and have a much more accurate knowledge of the traffic demand. For example, one destination might be shared by 100 TCP flows, while another contains only one flow. Leaving the outcome up to TCP results in a much fairer outcome than enforcing an equal amount of capacity per destination.

There are known problems with leaving fairness to endpoints, e.g., it is very easy to cheat by being less responsive to congestion \cite{briscoe2007flow}, but these problems also exist in current networks and are out of scope for this paper.

\section{Evaluation} 
\label{sec:evaluation}

\subsection{Implementation}

For the implementation of \tename, we chose Named Data Networking (NDN) \cite{zhang2014named}, since it has a mature codebase and built-in support for multipath forwarding at each hop. 
However, we do not use any NDN features other than those described earlier in this paper: a FIB with multiple nexthops per destination that can store some state (the split ratio and probing information) for each of these nexthops. 
Thus, \tename can be implemented on IP routers by adding only these features.

We evaluate \tename against related work, using the ns-3 network simulator \cite{henderson2008ns3} with the ndnSIM module \cite{afanasyev2012ndnsim}.

\subsection{Forwarding Adjustment \& Failure Resilience}

\begin{figure}
\centering
\includegraphics[width=.8\linewidth]{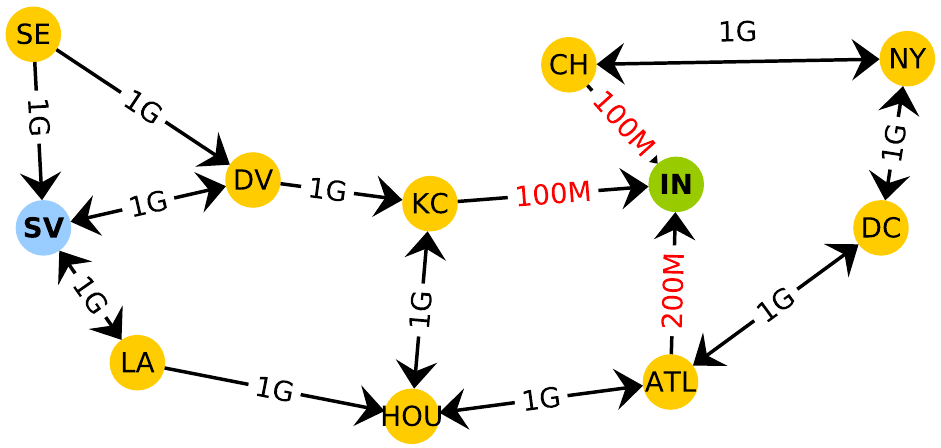}
\caption{Abilene Topology with routing for destination IN 
\vspace{\figspace} 
}
\label{fig:abilene}
\end{figure}

First, we use a small scenario to show the basic functionality of \tename over time. 
Specifically, we show its behavior for low traffic demand, high demand, and link failures.
We use the well-known Abilene topology (Figure \ref{fig:abilene}) with traffic from Sunnyvale (SV) to Indianapolis (IN). 
The arrows in Figure \ref{fig:abilene} show the viable nexthops, as computed by the multipath routing protocol.
We calculate the link propagation delays based on the geographical distance (obtained from the dataset of the Internet Topology Zoo \cite{internet_topology_zoo}) with a packet speed of 0.6c, which gives results closer to real-world RTTs than using the full speed of light.
The resulting one-way link delays range from 1.46 ms (CH, IN) to 12.3 ms (LA, HOU). 
These delays are also used as link weights, which determine the initial shortest path.
The link capacities (100 Mbps to 1 Gbps) are chosen so that the bottleneck will be at the three links adjacent to Indianapolis.

This scenario includes three distinct parts: 
\begin{itemize}
\item 0-50s: The application demand (\texttildelow 30 Mbps) is below the capacity of the shortest path.
\item 50-300s: The demand rises above the SP capacity.
\item 300-450: Link (ATL, IN) has failed.  
\end{itemize}

\begin{figure}
\centering
\includegraphics[width=\linewidth]{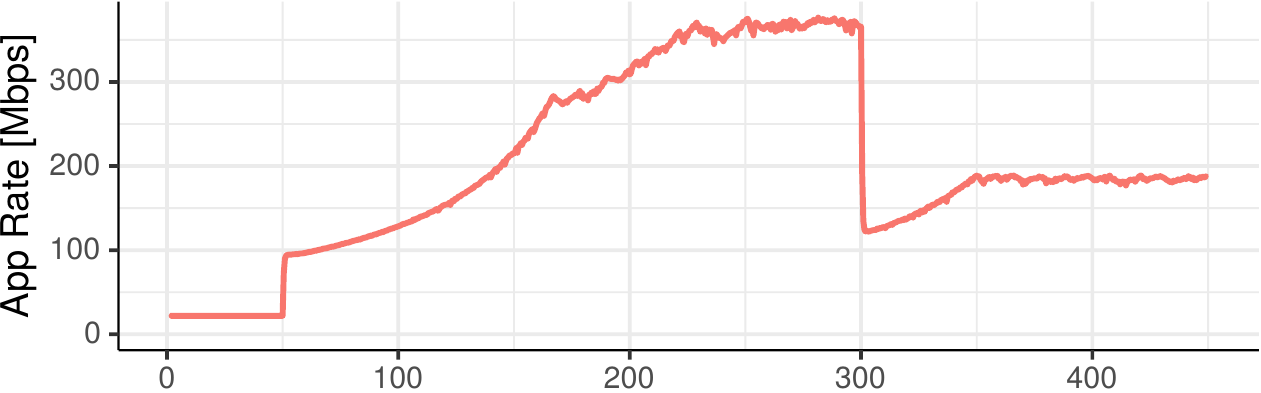}\\[.3em]
\includegraphics[width=\linewidth]{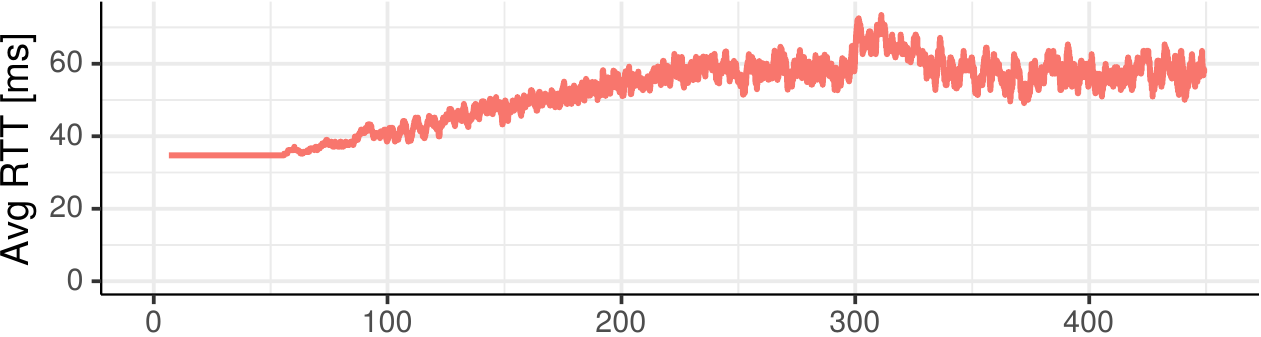}
\includegraphics[width=\linewidth]{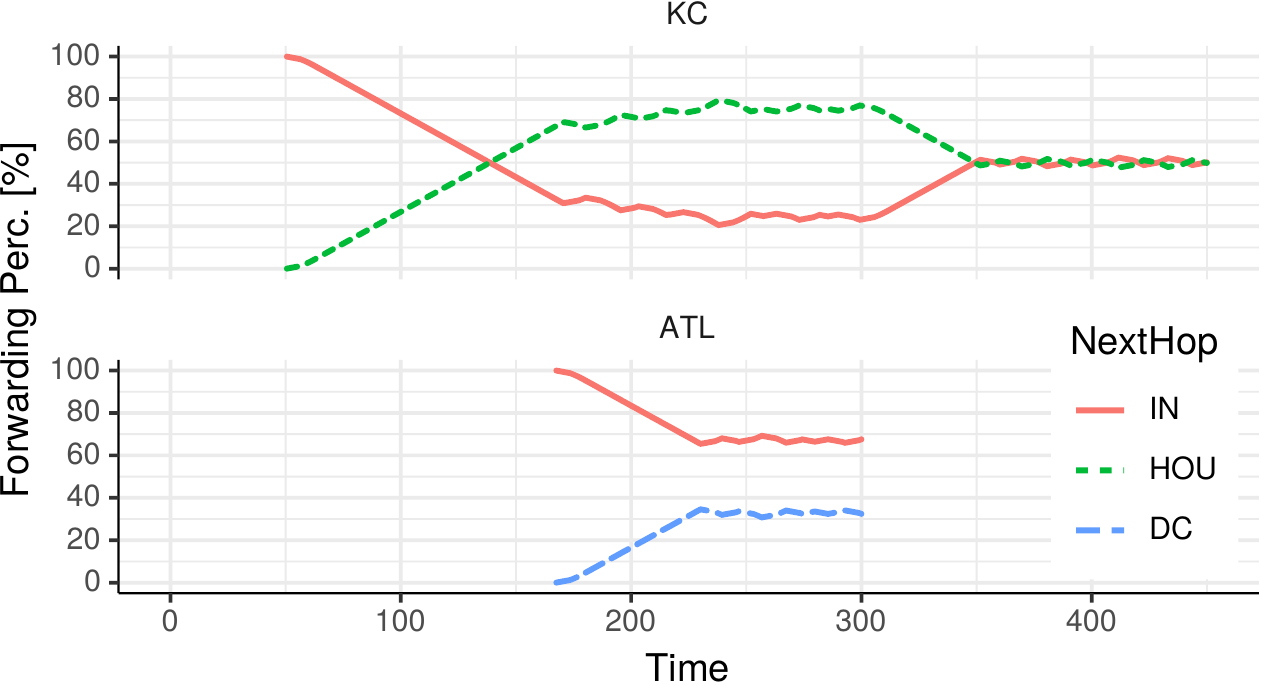}

\caption{Results for the Abilene Topology
\vspace{\figspace}
}
\label{fig:abilene_res}
\end{figure}

The results are shown in Figure \ref{fig:abilene_res}. First, as long as the demand is below link capacity, all packets stay on the shortest path SV$\rightarrow$DV$\rightarrow$KC$\rightarrow$IN. 

As soon as the demand exceeds capacity, router KC will see congestion marks from IN, start probing nexthop HOU, and then start shifting over traffic to a second path: KC$\rightarrow$HOU$\rightarrow$ATL$\rightarrow$IN until the KC's split ratio reaches about 66\% towards nexthop HOU (based on the difference in bottleneck capacity of the two paths). 
At that point, ATL will see congestion from IN, probe its neighbors, and start splitting traffic towards a third path ATL$\rightarrow$DC$\rightarrow$NY$\rightarrow$CH$\rightarrow$IN. 
The final split ratios will be KC $\Rightarrow$ \{IN: 25\%, HOU: 75\%\}; ATL $\Rightarrow$ \{IN: 66\%, DC: 33\%\}.
This is the optimal for this scenario, since the application throughput approaches the combined capacity of all three bottleneck links (400 Mbps). The average RTT increases corresponding to the longer paths that are used. 

At 300s, the link (ATL, IN) fails, and ATL immediately shifts all of its traffic to nexthop DC. Afterwards, KC will re-balance its forwarding split ratio to match the new optimum of \{IN: 50\%, HOU: 50\%\}.

\subsection{Traffic Split in Equilibrium}
\label{sec:ev_geant}

Next, we evaluate \tename's equilibrium traffic distribution for a larger, more diverse topology. 
Following our definition of latency-optimal routing, we need to evaluate two cases: 
1) When traffic demand stays below link capacity, \tename should only choose the shortest path. 
2) When traffic demand exceeds capacity, \tename should add longer paths, maximizing the aggregate throughput or the sum of user utilities.

The first case is obviously met, since traffic stays on the shortest path unless congestion marks are observed, implying that demand has exceeded link capacity. 
Hence, we focus our evaluation on the second case, and indeed use an \emph{infinite} demand between each source-destination pair.

\begin{figure}
\centering
\includegraphics[width=\linewidth]{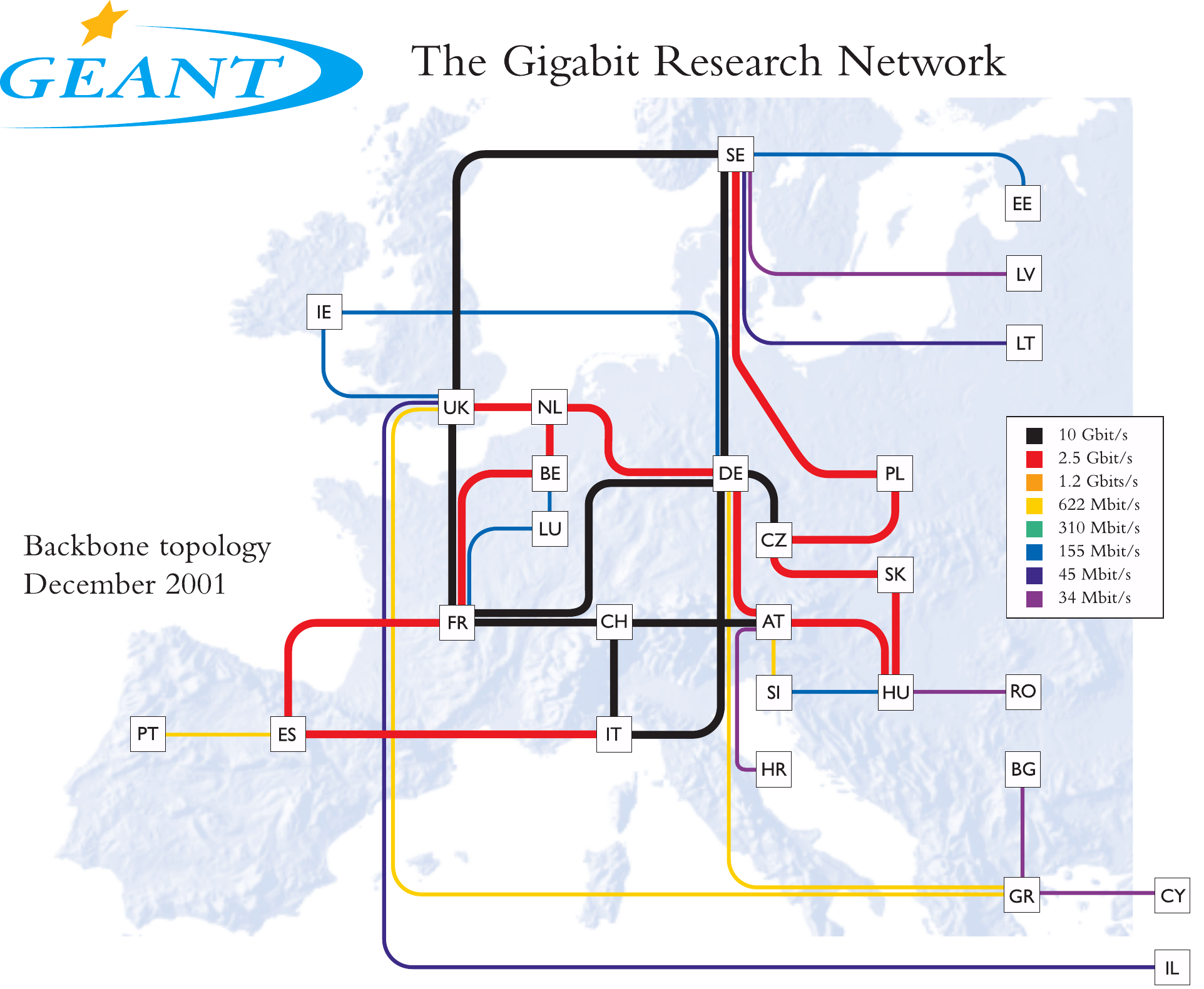}
\caption{Geant Backbone Topology 2001 \cite{geant_topo} 
\vspace{\figspace} 
}
\label{fig:geant}
\end{figure}

We use the Geant topology (Figure \ref{fig:geant}) since it is larger than the Abilene topology, has a higher average node degree, and also a large range of link capacities (from 34 Mbit/s to 10 Gbit/s). 
We calculate the link delays based on geographical distance (similar to above), which range from 0.90 ms (SK, HU) to 20.0 ms (IL, UK).
We compare \tename with related work from both IP and NDN multipath routing/traffic engineering:

\begin{itemize}
\item \textbf{SP:} Shortest path routing with link weights set to the propagation delay. 

\item \textbf{InvCap:} Shortest path routing with link weights set inversely to the link capacity. This is a simple and very common heuristic to distribute the link load.	

\item \textbf{PI:} Setting the forwarding split ratio in order to equalize the number of pending Interest packets (= content chunk requests) at each router. This is similar to the forwarding adaptation in \cite{carofiglio2016optimal}. 

\item \textbf{\tename:} Our algorithm, as described in this paper. 

\item \textbf{OPT:} The optimal result, obtained from solving the following optimization problem in \emph{Matlab}:
\begin{align}
\max_{\mathbf{x}, \mathbf{R}} \qquad 
& \sum_{i\in \mathcal{U}} U_i(x_i) - \gamma \sum_{l\in \mathcal{L}}D_{l}(f_{l})& & \label{eqObj} \\
\text{s.t.} \qquad 
&\mathbf{R}\mathbf{x} = \mathbf{f} \preceq \mathbf{c}, \quad \mathbf{x} \succeq \mathbf{0}, 
\end{align}
Where the utility $U_i(x) = -\frac{1}{x}$ is based on throughput $x$ for each application $i$, and $D_{l}(f_{l})$ represents the incurred link propagation delay (see details in Appendix \ref{sec:matlab}).
By choosing a small positive value for $\gamma$, the solution will maximize the utility with the shortest possible paths.

Note that this optimum is usually not achievable via loop-free hop-by-hop multipath routing (the optimal solution can include paths outside the restrictions imposed by ``downward paths'' \cite{source_selectable_2006} or LFID \cite{schneider2019hop}), and serves to show the theoretical limit of the other schemes. 
\end{itemize}

To compare the routing schemes, for each run we randomly distribute a number of source-destination pairs among the topology. We exclude the nodes that have only one link (BG, CY, IL, PT, HR, RO, LT, LV, EE), since this link is usually the bottleneck (capacities of 34 to 622 Mbps) and thus routing will not make a difference. 

We put a number of TCP Cubic-like applications (= 10) at each source node to retrieve packets from the producer application at the destination node. 
These apps will create infinite demand, that is, fill up the available link capacity. 
We run two scenarios (1 \& 5 src-dst pairs), with 20 runs each.
We chose the number of src-dst pairs quite low ($\leq$5), since with an increasing number, the optimal solution (and \tename) will approximate shortest path routing. 
If every node requests content with infinite demand, there is little use of routing on longer paths, since those paths will have little free capacity. 
In reality, it is unlikely that more than a few src-dst pairs will exceed their shortest path capacity. 
So you can think of the src-dst pairs here as modeling only those high-demand pairs.

We measure the total throughput, average completion time of a 10 Megabyte file, and average RTT over all used paths. 
Note that minimizing the average completion time (file size / throughput) is equivalent to maximizing the sum of user utility for utility function of $U(x) = -  \frac{1}{x}$, where x is the throughput. 
Thus, this utility function is also called \emph{minimum potential delay fairness} \cite{kunniyur2003minpot,floyd2008metrics_rfc}. 
We chose the average completion time as the main metric, since 1) it takes fairness into account (one starved application barely changes total throughput, but increases avg. completion time to infinity), and 2) is what TCP implicitly optimizes for \cite{kelly1998rate,low2003duality}. 
Hence, we use the same optimization goal for OPT.

\begin{figure}
\centering
\includegraphics[width=\linewidth]{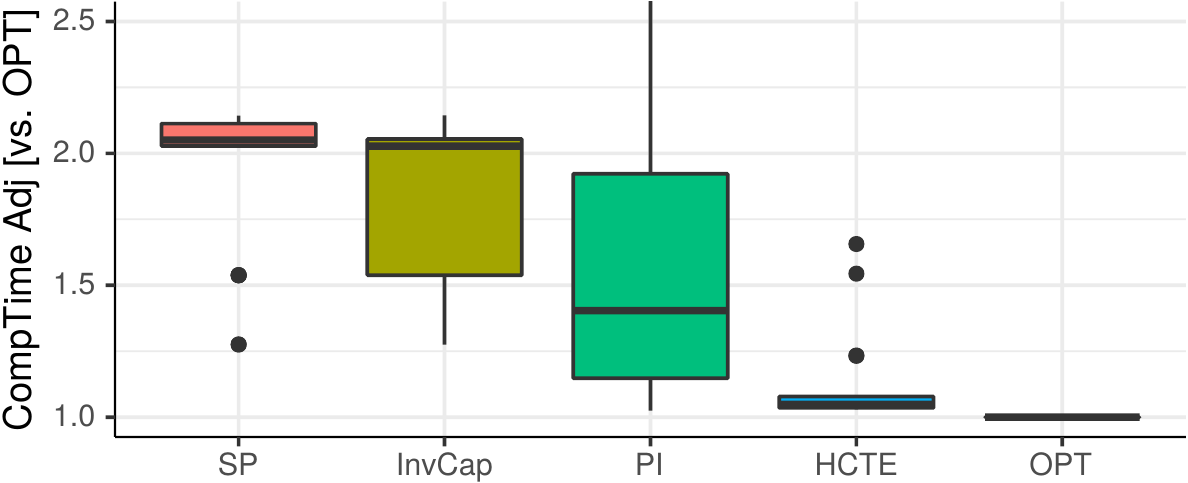}\\[.5em]
\includegraphics[width=\linewidth]{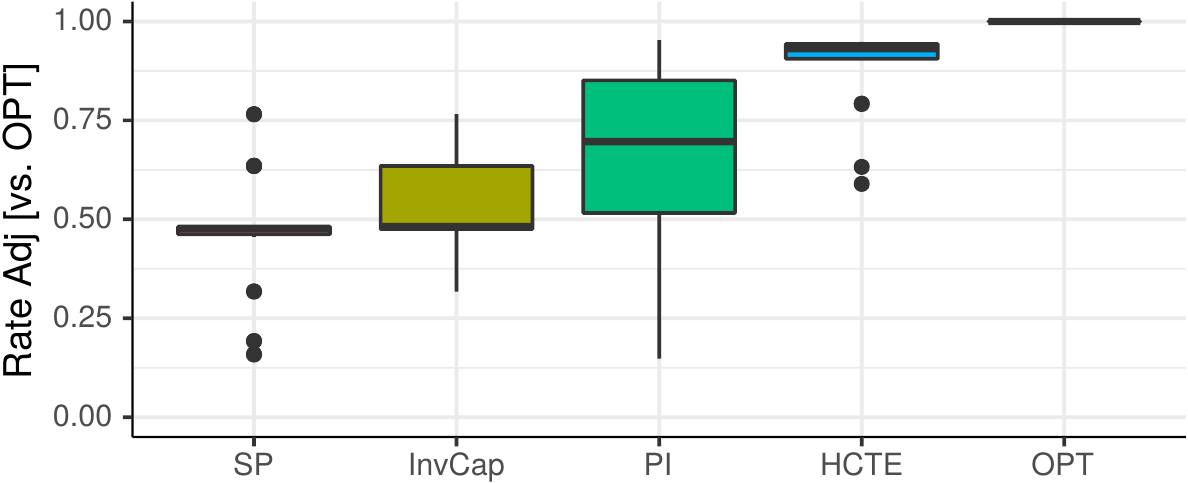}\\[.5em]
\includegraphics[width=\linewidth]{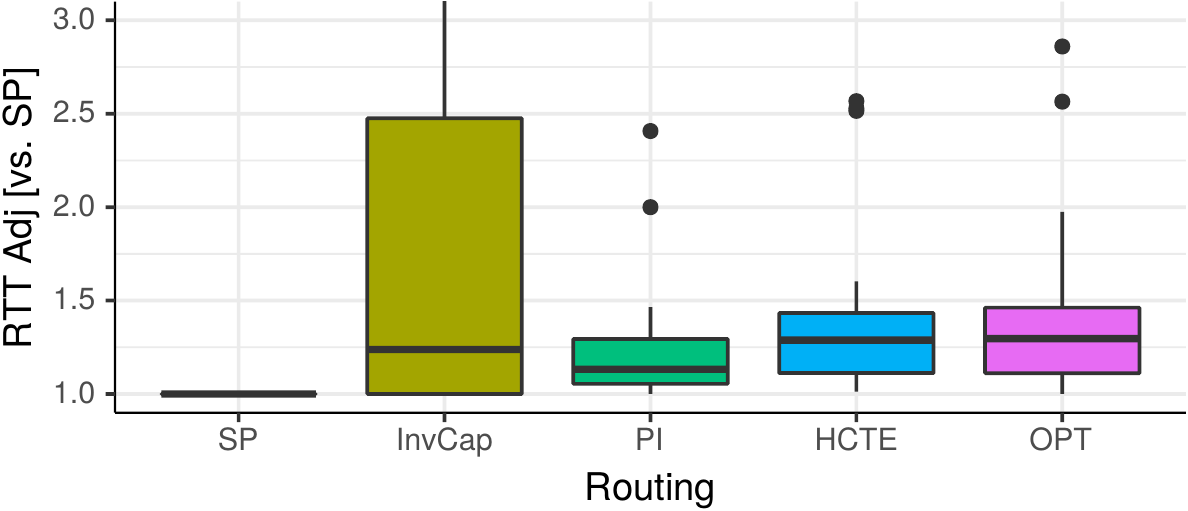}

\caption{Results for 1 randomly distr. src-dst pair (20 runs). \vspace{\figspace}
}

\label{fig:geant_res1}

\end{figure}
\begin{figure}
\centering
\includegraphics[width=\linewidth]{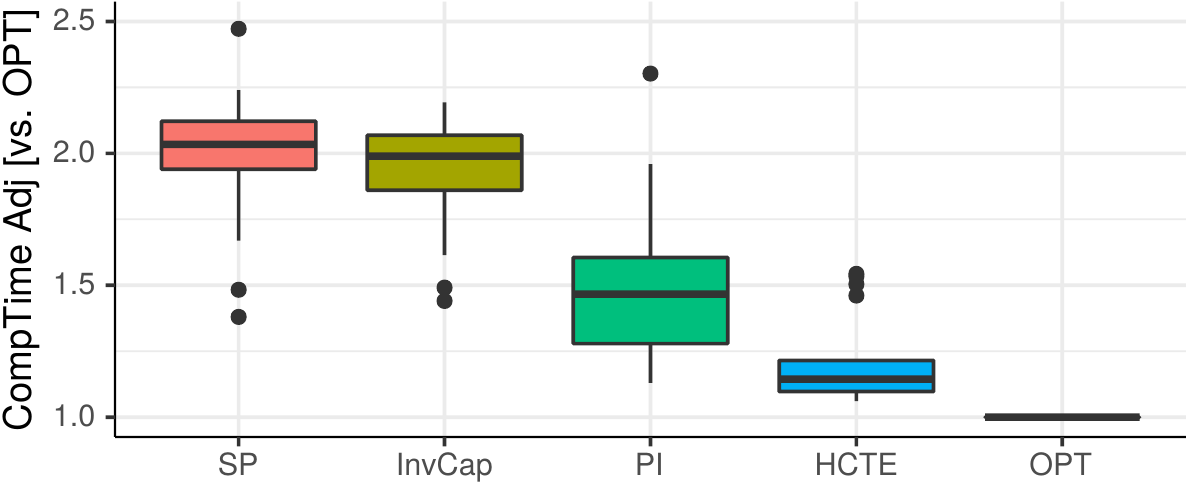}\\[.5em]
\includegraphics[width=\linewidth]{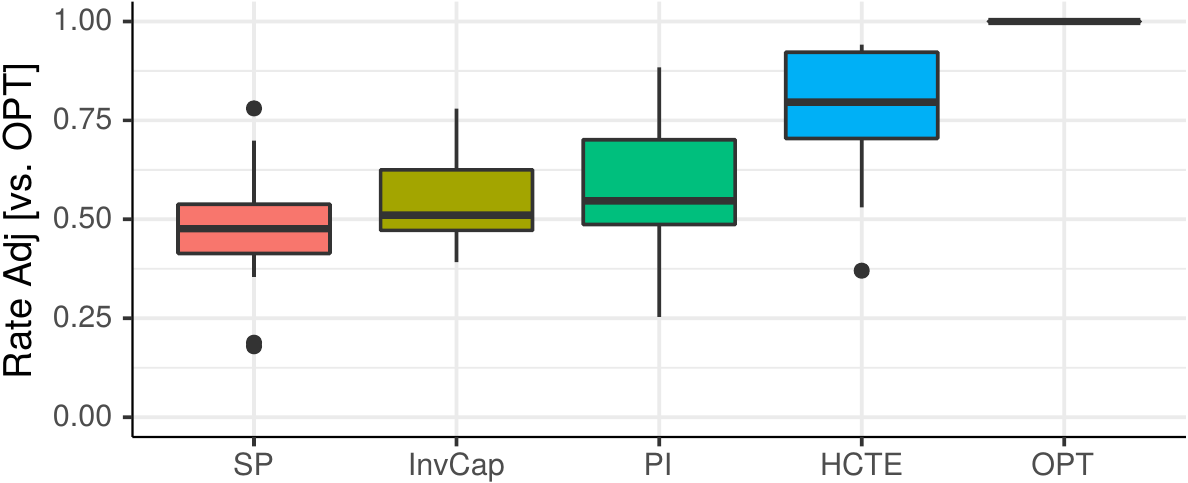}\\[.5em]
\includegraphics[width=\linewidth]{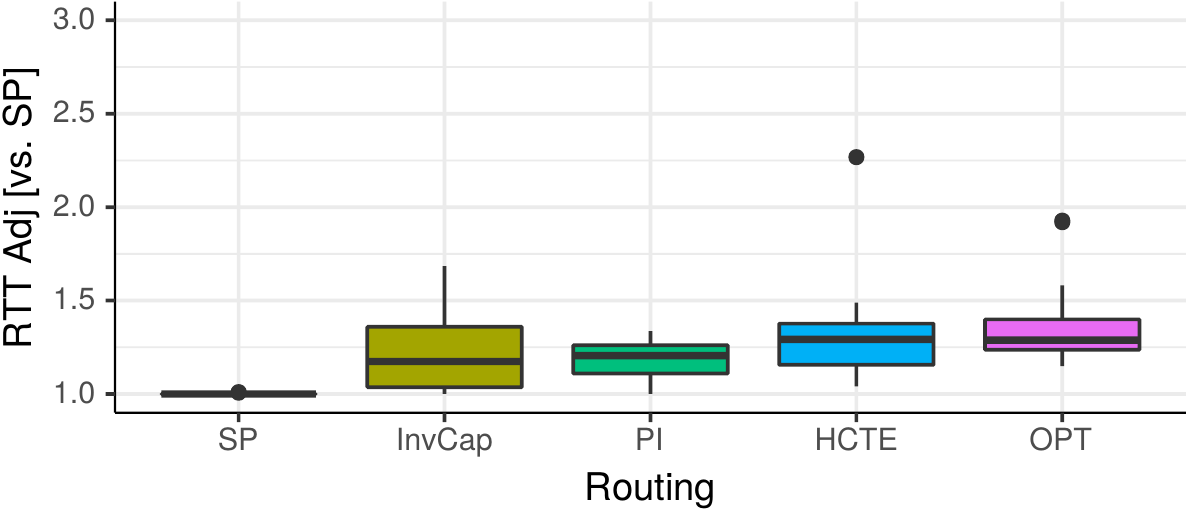}

\caption{Results for 5 randomly distr. src-dst pairs (20 runs). \vspace{\figspace}
}

\label{fig:geant_res2}

\end{figure}

The results in Figures \ref{fig:geant_res1} and \ref{fig:geant_res2} show a boxplot of 20 runs (median, hinges = 25th \& 75th percentiles, whiskers = 1.5*inter-quartile range), with the points showing outliers. 
Note that some of the outliers of the completion time (mostly SP and InvCap)  have been cut off from the top of the plot. 

It may be insightful to discuss the three outliers of \tename in Figure \ref{fig:geant_res1}. 
The one with the highest relative completion time (1.65x of OPT) is sending traffic from CH to UK. 
It will use the shortest path CH$\rightarrow$FR$\rightarrow$UK and a second slightly longer one CH$\rightarrow$AT$\rightarrow$DE$\rightarrow$NL$\rightarrow$UK (roughly 2.5x as long).
However, due to restrictions to get loop-free HBH routing, DE cannot use the nexthop SE (the path over SE$\rightarrow$UK has very high BW), since packets could take the path SE$\rightarrow$PL$\rightarrow$CZ$\rightarrow$DE, causing a loop. Also, if this path (CH$\rightarrow$AT$\rightarrow$DE$\rightarrow$SE$\rightarrow$UK) was available, it would be 5.28x longer than the shortest path.
The second highest outlier is traffic from NL to FR. NL can use the shortest path nexthop over UK or a slightly longer (1.6x) path over BE. 
However, NL cannot choose the path over DE (3.33x longer), since it may cause the loop DE$\rightarrow$IE$\rightarrow$UK$\rightarrow$NL.
The third outlier is similarly caused by restrictions in loop-free HBH routing.
These outliers show that sometimes \tename diverges from the optimal.
However, they are rare and only very high-delay paths (at least 3x longer than shortest path) are omitted. 

In general, for both scenarios (1 and 5 src-dst pairs), \tename clearly outcompetes the related work in terms of completion time. The median completion time (over 20 runs) is only 6\% or respectively 14\% higher than the optimal. 
Note that there are multiple overheads in the ndnSIM evaluation that don't exist in the Matlab calculation of the optimal result: 
the size of packet headers, probing traffic, and fluctuations in both TCP rate adjustment and router split ratio adjustment. 

The total throughput of \tename is less close to the optimal (median of 93.1\% and 79.6\% respectively), but still higher than related work. For the average RTT, the optimal solution is higher than either PI or \tename, since it uses some longer paths that are excluded from Hop-by-hop routing.

\section{Related Work}
\label{sec:related_work}

One kind of related work is Multipath TCP (MPTCP) \cite{wischik2011mptcp}, a \emph{transport layer} protocol which runs on end devices to simultaneously multiple access networks, e.g., a smart phone connecting to both WiFi and a cellular network. 
MPTCP only determines the first hop and routing is left to the corresponding access networks.
Thus, MPTCP paths often show significant overlap.
In contrast, \tename works at the \emph{network layer} and does influence routing decisions.
One interesting aspect is how both schemes deal with \emph{shared bottlenecks.}
In MPTCP, two subflows may share the same bottleneck link and hence would achieve twice the throughput as one competing TCP flow.
MPTCP solves this issue by making each subflow proportionally less aggressive than TCP.
In \tename, we instead avoid the problem of shared bottlenecks by rerouting only at those bottleneck links. 
Since traffic is only split directly adjacent to (or as close a possible to) a congested link (see Section \ref{sec:design}), rerouted flows will never return to the same bottleneck link.

Two other related works are DEFT and PEFT \cite{xu2007deft,xu2011peft}, which share our approach of using a link congestion price and hop-by-hop forwarding. 
The main difference is that DEFT/PEFT uses a fixed exponential split ratio at routers and manipulates link weights.
\tename uses fixed link weights (at the propagation delay) and manipulates the split ratio. 
We believe that our approach is less vulnerable to the global side-effects of tuning link-weights. 
Moreover, the fixed exponential split ratio means that traffic will be split even if the shortest path has sufficient capacity,
violating our definition of ``latency-optimal routing'' (see Section \ref{sec:optimal}).

Some related works are close to the generic source-routing discussed in Section \ref{sec:history_source}: TeXCP, MATE, DATE, and COPE.
TeXCP \cite{texcp2005walking} is a distributed (among ingress routers), online traffic engineering scheme that uses MPLS to balance the load among multiple end-to-end paths.
It considers potential traffic oscillations when multiple ingress routers adjust to congestion on a single link. 
TeXCP solves this by using explicit feedback to assure the traffic increase on both paths will not overshoot the link capacity.
Our solution to the oscillation problem is quite different: Only a single entity, the router closest to the congestion, will adjust the split ratio. 
This reduces oscillations, since one of the main issues is multiple independent entities making the same decision at the same time. 
Moreover, the router adjacent to congestion has a much quicker feedback loop than edge routers, which may further reduce oscillations.
TeXCP mostly has the same issues as the generic source-routing (Section \ref{sec:history_source}).
It uses MinMax routing as optimization goal, which the authors admit ``may sometimes increase the delay'' \cite{texcp2005walking}.
Their solution is to have the operator manually limit the possible path choices: ``Clearly the ISP should not pick a path that goes to LA via Europe'' \cite{texcp2005walking}.
In contrast, our traffic engineering solution requires no such intuition or common sense from the operator. Paths will be the shortest possible by default, and longer paths are only considered when demand exceeds the capacity of shorter ones.

MATE, DATE, and COPE are similar to TeXCP in many regards, so we describe only the differences.
MATE \cite{mate_elwalid2001} is different in that 1) it minimizes a cost function based on a model of queuing delay rather than minimizing maximum link utilization. 2) MATE assumes that ingress nodes have instantaneous knowledge of the whole network state, while TeXCP is fully distributed \cite{texcp2005walking}.
COPE \cite{wang2006cope} provides a worst-case performance guarantee under all traffic demands.  
DATE \cite{he2007date} explicitly considers congestion prices and tries to maximize user utility (rather than minimizing link utilization or cost), which is similar to our approach. 

Lastly, we discuss two works that are more closely related to \tename.
HALO \cite{michael2013optimal,michael2015halo} (Hop-by-Hop Adaptive Link-state Optimal Routing) also splits traffic at each router and is ``adaptive'', i.e., does not require to estimate the traffic demand matrix.
However, like the work discussed above, the optimization goal is still MinMax routing, which does not account for a path's propagation delay, thus often sends packets on paths longer than necessary. 

Most similar to our work is the work by Kvalbein et al. called ``Routing Homeostasis'' (RH) \cite{kvalbein2009multipath}.
RH uses a distributed hop-by-hop approach, where individual routers decide on their nexthops.
It has the same optimality goal as \tename, of preferring low-latency paths, and it doesn't relying on estimation of traffic matrices.
The main differences are that RH works completely local without a way of probing path congestion: routers adjust the split ratio based only on the utilization of their adjacent links, without regard to the path further down in the network. 
This can lead to congested links further along the path which are not considered, but which would be detected by our path probing mechanism.
Kvalbein et al. also currently do not have any packet-level simulator implementation of their scheme, which made it hard to compare against.

\section{Conclusions}

We presented a renewed case for Hop-by-Hop Traffic Engineering, which could be a viable alternative to current source routing traffic engineering. 
It remains to be seen in which environments the simpler, distributed design is preferable over a more powerful but complex, centralized approach.

\appendix
\subsection{Optimization Problem}
\label{sec:matlab}

We consider a multi-commodity flow (MCF) problem that is well-known in the literature \cite{wang2005cross,lin2006utility}, where a network is modeled as a set of directional links, denoted by $\mathcal{L}$, with finite capacities $\boldsymbol{c} = (c_l, l \in \mathcal{L})$ and propagation delays $\boldsymbol{d} = (d_l, l \in \mathcal{L})$, shared by a set of source-destination pairs $\mathcal{U}$. Let  $\mathbf{x} = (x_i, i\in \mathcal{U})$ denote the flow vector and $\mathbf{R}=[R_{li}]$  the routing matrix that obeys the flow conservation rules, i.e., $x_i$ is the data flow of pair $i\in \mathcal{U}$ and $R_{li}$ the fraction of $i$'s flow along link $l$. 
The objective is a joint user performance maximization and network cost minimization:
\begin{align}
\max_{\mathbf{x}, \mathbf{R}} \qquad 
& \sum_{i\in \mathcal{U}} U_i(x_i) - \gamma \sum_{l\in \mathcal{L}}D_{l}(f_{l})\\
\text{s.t.} \qquad 
&\mathbf{R}\mathbf{x} = \mathbf{f} \preceq \mathbf{c}, \quad \mathbf{x} \succeq \mathbf{0}, 
\end{align}
Where the network cost is represented by the network-wide flow delay cost  $D_{l}(f_{l}) = d_{l} * f_{l}$, where $f_{l}$ is the flow through link $l$. 
Assuming that the user utilities $U_i:\mathbb{R}_+\to \mathbb{R}$ are concave and the link cost functions $D_{l}:[0,c_{l})\to \mathbb{R}$ are convex, this is a MCF problem with linear constraints and convex objective, and thus can be solved efficiently by various numerical packages.

We chose a utility function inversely proportional to the throughput, since it is implicitly  achieved by TCP congestion control \cite{kelly1998rate,low2003duality}:
$ U_i(x) = -\frac{1}{x}, \quad \forall i\in \mathcal{U}$

\bibliographystyle{abbrv}
\bibliography{content/bib}

\end{document}